\documentstyle[12pt]{article}
\begin{document}
\baselineskip 3.9ex
\def\be{\begin{equation}}
\def\ee{\end{equation}}
\def\ba{\begin{array}{l}}
\def\ea{\end{array}}
\def\bea{\begin{eqnarray}}
\def\eea{\end{eqnarray}}
\def\no#1{{\tt   hep-th#1}}
\def\eq#1{(\ref{#1})}
\def\pgap{\vspace{1.5ex}}
\def\ggap{\vspace{10ex}}
\def\gap{\vspace{3ex}}
\def\del{\partial}
\def\o{{\cal O}}
\def\z{{\vec z}}
\def\re#1{{\bf #1}}
\def\av#1{{\langle  #1 \rangle}}
\def\S{{\cal S}}

\renewcommand\arraystretch{1.5}

\begin{flushright}
TIFR/TH/99-15\\
April 1999\\
hep-th/yymmdd
\end{flushright}
\begin{center}
\vspace{2 ex}
{\large{\bf Anti-de Sitter gravity associated with the supergroup
$SU(1,1|2)\times SU(1,1|2)$}}\\
\vspace{3 ex}
Justin R. David$^*$   \\
{\sl Department of Theoretical Physics} \\
{\sl Tata Institute of Fundamental Research,} \\
{\sl Homi Bhabha Road, Mumbai 400 005, INDIA.} \\
\vspace{10 ex}
\pretolerance=1000000
\bf ABSTRACT\\
\end{center}
\vspace{1 ex}

We construct the anti-de Sitter supergravity in three dimensions
associated with the supergroup $SU(1,1|2)\times SU(1,1|2)$. 
The field content and the action are inferred using the fact that $AdS$
supergravity theories in three dimensions are Chern-Simons theories.

\vfill
\hrule
\vspace{0.5 ex}
\leftline{$^*$ justin@theory.tifr.res.in}
\clearpage

\section{Introduction. }

Maldacena's $AdS/CFT$ correspondence 
\cite{MAL} has spurred interest in Anti-de
Sitter supergravity theories. A particularly interesting case is the
$AdS_3/CFT_2$ correspondence \cite{MALSTR,MAR,BOER,SEIKUTGIV,DMW}. 
In this case string theory on $AdS_3\times S^3\times M_4$ 
with $RR$ flux is
conjectured to be dual to the conformal theory of the Higgs branch of
the D1-D5-brane system, where $M_4$ can be $T^4$ or $K_3$. 
It is relevant to understand the
corresponding supergravity on $AdS_3$ for the investigation of this
conjecture. On examining the symmetries of the near horizon geometry
of the D1-D5 system we find that the 
specific $AdS_3$ supergravity is the one
associated with the supergroup ${\cal G}= SU(1,1|2)\times SU(1,1|2)$.
In this letter we will construct the pure 
( i.e. no coupling to matter )
anti-de Sitter supergravity associated
with the above supergroup. 

Construction of this supergravity would be useful in furthering our
understanding of the boundary conformal field theory. It is well known
that gravity in $AdS_3$ is a Chern-Simons theory with the gauge group
$SL(2,R)\times SL(2,R)$ and it can be reformulated as a Liouville
theory at the boundary \cite{CHD}. 
A Liouville theory corresponding to
pure $AdS$ supergravity associated with the supergroup ${\cal G}$
is likely to help us
understand the space time conformal field theory constructed in 
\cite{SEIKUTGIV} which is also a Liouville theory. Secondly,
construction of this supergravity can help us study  backgrounds
in $AdS_3$ other than the pure $AdS_3$ and the $BTZ$ black hole, 
which are the
backgrounds being studied presently. These other backgrounds can help us
understand certain properties of the boundary CFT. 
In fact the motivation for construction of this $AdS$ supergravity
arose from the problem of supersymmetrically 
embedding the conical spacetimes of \cite{DES} in this supergravity.
These spaces provide a one parameter family of metrics
which interpolate between pure $AdS_3$ and the zero mass $BTZ$ black
hole mimicking the spectral flow of the CFT \cite{DMVW}. 

To construct the  $AdS$ supergravity on ${\cal G}$ we
use the fact that this theory is a Chern-Simons theory associated with
the supergroup ${\cal G}$. That is, 
the action can be written formally
as
\be
\int_{M_3} tr (\Gamma d\Gamma + \frac{2}{3} \Gamma^3)
\ee
where $\Gamma$ is the connection one-form for the supergroup and $M_3$
is the  three-dimensional bosonic submanifold of the 
supergroup ${\cal G}$ with local coordinates $x^\mu$. 
$\Gamma$ does not depend on other coordinates of the group manifold
\cite{ACHTOW}.
Using this one can fix the field content using the
generators of the supergroup. The equations of motion of the
$AdS$ supergravity on ${\cal G}$ will be the 
Maurer-Cartan equations of the supergroup
which can be written down from the corresponding superalgebra.  

This letter is organized as follows. In Sec. 2 we write down the
superalgebra on ${\cal G}$ in a form suitable to extract
the field content of the $AdS$ supergravity and list the Maurer-Cartan
equations for the supergroup. In Sec. 3 we present the action for the 
$AdS$ supergravity on ${\cal G}$. Then we list down the
supersymmetry transformation laws and show that the action is invariant
under them. In Sec. 4 we state our conclusions.

\section{The  $SU(1,1|2) \times SU(1,1|2)$ superalgebra. }

The $SU(1,1|2) \times SU(1,1|2)$ super algebra is the global part of
the small ${\cal N} =(4,4)$ super conformal theory (see for e.g.
\cite{BOER}). The bosonic 
generators consist of $L_{-}, L_{+}, L_0$ which form the $SL(2,R)$
part of the algebra and $T^i , T'^i $ which are 
the global $SU(2)$ currents.
The supercharges consist of $G_{-1/2}^{\alpha}, G_{1/2}^{\alpha}, 
G'^{\alpha}_{-1/2}, G'^{\alpha}_{1/2} $ which
transform as fundamentals under the global $SU(2)$ currents. We can
organize the $SL(2,R)\times SL(2,R)$ generators into the generators of
Lorentz transformations $M_{ab}$ and translations $P_a$ of $SO(2,2)$,
the isometry group of $AdS_3$.
The supercharges of the each of the $SU(1,1|2)$ transform as a
Dirac fermion in $AdS_3$ with an internal $SU(2)$ index. Then the
$SU(1,1|2)\times SU(1,1|2)$ super algebra is given by the following
(anti) commutation relations. 

\bea
[P_a , P_b ] = -4 m^2 M_{a b}      \; &\;& \;
[P_a , M_{bc}] = - \eta_{ab} P_{c} + \eta_{ac} P_{b}   \\  \nonumber
[M_{ab}, M_{cd}] &=& - \eta_{ad} M_{bc} - \eta_{bc} M_{ad}
+ \eta_{ac} M_{bd} + \eta_{bd} M_{ac}  \\  \nonumber
[T^{i}, T^{j} ] = 4 i \epsilon_{ijk} T^{k}  \; &\;& \;
[T'^{i}, T'^ {j} ] = 4 i \epsilon_{ijk} T'^{k}
\\  \nonumber
[P_a, G] = m \gamma_a G   \; &\;& \;
[M_{ab}, G] = \frac{\gamma_{ab}}{2} G \\  \nonumber
[P_{a}, G'] = -m \gamma_{a} G'  \; &\;& \;
[M_{ab}, G'] = \frac{\gamma_{ab}}{2} G' \\ \nonumber
[T^{i}, G] = -2 \sigma^{i} G  \; &\;& \;
[T^{i}, G'] = -2 \sigma^{i} G'  \\   \nonumber
\{ G^{\alpha}, G^{\dagger \beta} \} 
&=& \left[\delta^{\alpha \beta} 
(-\frac{P_a \gamma^a}{4} +m \frac{M_{cd}\gamma^{cd}}{4} )
-m \frac{ \sigma^i_{\alpha\beta} 
T^{i}}{4} \right]\gamma_{0}    \\  \nonumber
\{ G'^{\alpha}, G'^{\dagger \beta} \}
&=& \left[ \delta^{\alpha \beta} 
(- \frac{P_a \gamma^a}{4} - m \frac{ M_{cd}\gamma^{cd} }{4} )
+m \frac{ \sigma^i_{\alpha\beta} 
T'^{i} } {4} \right] \gamma_{0}    \\  \nonumber
\eea

Our conventions are as follows. $a,b,c \ldots$ denote the tangent
space indices of the $AdS$ space. The tangent space metric has
signature $(-1,1,1)$. $i,j,k$ take values from $1$ to $3$
corresponding to the generators of $SU(2)$.
The supercharges $G$ and $G'$ are Dirac fermions in $AdS$ space
which transform in the fundamental of the two $SU(2)$'s respectively.
$1/m$ refers to the radius of $AdS_3$ which is related to 
the cosmological constant.
The $\gamma$ matrices are given by
\be
\gamma_0 = \left( \begin{array} {cc} 
0 &1  \\ -1 & 0  \end{array} \right), \;\;
\gamma_1 = \left( \begin{array}{cc} 
0 & 1 \\ 1 & 0 \end{array} \right), \;\;
\gamma_2 = \left( \begin{array}{cc} 
1 & 0 \\ 0 & -1 \end{array} \right)
\ee
$\sigma^i$ denotes the Pauli matrices and $\alpha , \beta $  
denotes
the $SU(2)$ indices. We define
$ \gamma_{ab}=(1/2)[\gamma_a, \gamma_b] $. The following
formulae are useful
\be
\gamma_{ab}= \epsilon_{abc}\gamma^c , \;\;\; \epsilon_{123}=1,
\;\;\; \epsilon_{abc}\epsilon^{abc}= -3!
\ee

To write the Maurer-Cartan equations we introduce the gauge field
one-forms
\bea
\omega_{ab}= dx^\mu \omega_{ab\mu},  \; &\;& \; e_a= dx^\mu e_{a\mu} \\
\nonumber
\psi^\alpha = dx^\mu \psi_\mu^\alpha  , \; & \;& \;
\psi^{'\alpha} = dx^\mu \psi_\mu^{'\alpha}   \\  \nonumber
A^i=dx^\mu A_\mu^i  ,    \; &\;& \;
A'^i=dx^\mu A'^i_\mu  
\eea
associated with the generators $M_{ab}, P_{a}, G^\alpha , G'^\alpha ,
T^i , T'^i $ respectively. In other words
\be
\Gamma = \omega_{ab}M^{ab} + e_a P^a + \bar{\psi}G + \bar{G}\psi +
  \bar{\psi'}G' + \bar{G}'\psi' + A^i T^i + A'^i T'^i
\ee
Thus the field content of the
$SU(1,1|2)\times SU(1,1|2)$ supergravity consists of a vielbein
$e_{a\mu},$ two $SU(2)$ gauge fields $A_\mu^i ,\,  A'^i_\mu $ and 
two gravitini $\psi_\mu^{\alpha},\,  \psi'^{\alpha}_\mu$ transforming
as the fundamental of the two $SU(2)$'s. The gravitini are Dirac
fermions. $\omega_{ab\mu}$ is the spin connection. 
The Maurer-Cartan equations for the one-forms turn out to be
\bea
\label{Maucar}
de_a &=& -\eta^{bc} e_b \omega_{ca} + \frac{ \bar{\psi}  \gamma_a
\psi }{4} + \frac{ \bar{\psi' } \gamma_a \psi'}{4}  \\ \nonumber
d\omega_{ab} &=& -4 m^2 e_a  e_b - \eta^{cd}\omega_{ac}
\omega_{db} +m \frac{ \bar{\psi} \gamma_{ab} \psi } {2} 
-m \frac{ \bar{\psi' \gamma_{ab}} \psi' }{2}  \\  \nonumber
d\psi &=& m e_a\gamma^a \psi - 
\frac{\omega_{ab}\gamma^{ab} \psi }{4} + 2A^i\sigma^i \psi \\
\nonumber
d\psi' &=& - m e_a\gamma^a \psi'
- \frac{\omega_{ab}\gamma^{ab} \psi' }{4}  +2A'^i\sigma^i \psi' \\
\nonumber
dA^i &=& 2i \epsilon_{ijk} A^j A^k -
m \frac{ \bar{\psi}\sigma^i \psi  } {4} \\   \nonumber
dA'^i &=& 2i \epsilon_{ijk} A'^j A'^k +
m \frac{ \bar{\psi'}\sigma^i \psi' } {4} \\   \nonumber
\eea
Here $\bar{\psi}$ is defined as $\psi^{\dagger}\gamma_0$.

\section{ The Action. }

The Maurer-Cartan equations of the supergroup $SU(1,1|2)\times
SU(1,1|2)$ written above give the equations of motion of the
supegravity we require. These equations can be obtained as the
Euler-Lagrange equations of the following action.
\bea
\label{action}
S &=&\!
\int d^3 x  \left[ \right. \frac{eR}{2} + 4m^2e  \\  \nonumber
-\frac{\epsilon^{\mu\nu\rho} }{2} \bar{\psi}_{\mu} {\cal D}_{ \nu}
\psi_{\rho} \!\!&-&\!\! \frac{2\epsilon^{\mu\nu\rho} } { m } ( A_{\mu}^i
\del_{\nu} A_{\rho}^i - \frac{ 4i\epsilon_{ijk} }{3} A_{\mu}^i
A_{\nu}^j A_{\rho}^k ) \\  \nonumber
-\frac{\epsilon^{\mu\nu\rho} }{2} \bar{\psi'}_{\mu} {\cal D'}_{ \nu}
\psi'_{\rho} \!\!&+&\!\! \frac{2\epsilon^{\mu\nu\rho}} { m }( A'^i_{\mu}
\del_{\nu} A'^i_{\rho} - \frac{ 4i\epsilon_{ijk} }{3} A'^i_{\mu}
A'^j_{\nu} A'^k_{\rho})  \left. \right] \\  \nonumber
\eea
Where ${\cal D}_{\nu }  = \del_{\nu} + \frac{\omega_{ab
\nu }\gamma^{ab}  }{4} - m e_{a \nu}\gamma^{a} -2 A_{ \nu}^i\sigma^i $
and 
${\cal D'}_{\nu }  = \del_{\nu} + \frac{\omega_{ab
\nu }\gamma^{ab} }{4} + m e_{a \nu}\gamma^{a} -2 A'^i_{ \nu}\sigma^i
$.The above Lagrangian reduces to
that of \cite{ISQTOW} if the gauge group is $U(1)$. This makes it easy
to generalize the supersymmetry transformations for the non-abelian
case. The first equation in (\ref{Maucar})    is the following 
constraint which defines $\omega_{ab\mu}$
\be
\del_{[\mu} e_{a \nu]} + \omega_{ab[\mu}e^{b}_{\nu]}=
\frac{1}{4}(\bar{\psi}_{[\mu}\gamma_a \psi_{\nu]}+
\bar{\psi'}_{[\mu}\gamma_a \psi'_{\nu]})
\ee
This is also the equation of motion obtained by variation of the
action in (\ref{action}) by 
$\omega_{ab\mu}$ treating the spin connection and the
vielbein as independent fields. Thus we use the ``1.5 order formalism
'' in verifying the invariance of the action under supersymmetry
transformations.
The supersymmetry transformations under which the action is invariant
are
\bea
\delta e^a_{\mu} &=& \frac{1}{4}\bar{\epsilon}\gamma^a \psi_{\mu}
+\frac{1}{4}\bar{\epsilon'}\gamma^a \psi'_{\mu} + cc  \\ \nonumber
\delta \psi_{\mu} &=& {\cal D}_{\mu}\epsilon \\ \nonumber
\delta \psi'_{\mu} &=& {\cal D'}_{\mu}\epsilon' \\ \nonumber
\delta A^i_{\mu} &=& -\frac{m}{4} \bar{\epsilon}\sigma^i \psi + cc \\
\nonumber
\delta A'^i_{\mu} &=& \frac{m}{4} \bar{\epsilon'} \sigma^i \psi' + cc 
\eea
here $cc$ denotes complex conjugate.
To verify these supersymmetry transformations it is helpful to write
the curvature and the cosmological constant term in the 
action (\ref{action}) as
\be
\int d^3x \left(
-\frac{1}{2} e_{a\rho}\epsilon^{\rho\mu\nu}\epsilon^{abc}
(\, \del_{[\mu}\omega_{bc\nu]} + \omega_{bd[\mu}\omega^d_{c\nu]}\, )
-\frac{2m^2}{3}
\epsilon^{\mu\nu\rho}\epsilon^{abc}e_{a\mu}e_{b\nu}e_{c\rho} \right)
\ee
In course of the verification one  arrives at the following
typical four-fermion term
\be
\frac{m}{4} \left[
-\epsilon^{\mu\nu\rho}(\, \bar{\psi}_{\mu}\gamma_a \psi_\nu
\, ) (\, \bar{\epsilon}\gamma^a \psi_\rho \, )
+ \epsilon^{\mu\nu\rho} (\, \bar{\psi}_\mu \sigma^i \psi_\nu \, ) 
(\,  \bar{\epsilon}\sigma^i \psi_\rho \, ) \right]
\ee
It can be shown that such terms cancel by means of a Fierz indentity
for non-commuting fermions which is
\be
(\,\bar{\lambda} \psi \, )(\, \bar{\xi} \chi\, ) = -\frac{1}{2}
\left[\,  (\,\bar{\lambda}\chi\, )(\, \bar{\xi}\psi\, ) + 
(\, \bar{\lambda}\gamma^a\chi \, )
(\, \bar{\xi}\gamma_a \psi\, )\,\right]
\ee
The additional four-fermion cross-terms simply cancel. 

\section{Conclusions.}

In this letter we have constructed the Anti-de Sitter supergravity
associated with the supergroup $SU(1,1|2)\times SU(1,1|2)$. To
construct it we used the fact that the action can be written as the
integral of the Chern-Simons three-form associated with the
supergroup. It is also important to note that
this procedure of constructing $AdS$ supergravity can be generalized
to other three-dimensional $AdS$ supergroups for which the
corresponding supergravity has not yet been constructed. 
It would be interesting to couple this pure supergravity
to matter. That such a coupling should 
exist is clear from the fact that
the supergravity describing the near horizon geometry of the D1-D5
system contains matter as well. This would give
further understanding of the Maldacena correspondence for this case.

\section{Acknowledgments.}

The author wishes to acknowledge fruitful discussions with  
Gautam Mandal and Spenta Wadia 
which provided the motivation for construction of the action. 
He also gratefully acknowledges discussions with S. Vaidya who
participated in the early phase of this project.

\noindent
{\bf Note added:} After this work was completed we received \cite{TAN}
which overlaps with some portion of this letter.

\end{document}